\documentclass[iop]{emulateapj}

\newcommand{\icarus} {Icarus}

\newcommand{\Rp}{\ensuremath{R_{\rm p}}}
\newcommand{\rhop}{\ensuremath{\rho_{\rm p}}}
\newcommand{\Mp}{\ensuremath{M_{\rm p}}}
\newcommand{\gcc}{\ensuremath{\,{\rm g\,cm}^{-3}}}
\newcommand{\rroche}{\ensuremath{R_{\rm Roche}}}
\newcommand{\rL}{\ensuremath{R_{\rm L}}}
\newcommand{\RH}{\ensuremath{R_{\rm H}}}

\begin{document}

\title{Warm Saturns: On the Nature of Rings around Extrasolar Planets that
  Reside Inside the Ice Line}

\author{Hilke E. Schlichting\altaffilmark{1,2,3} \& Philip
Chang\altaffilmark{4} } \altaffiltext{1} {UCLA, Department of Earth and Space
Science, 595 Charles E.  Young Drive East, Los Angeles, CA 90095, email:
hilke@ucla.edu}
\altaffiltext{2} {California Institute of
Technology, MC 130-33, Pasadena, CA 91125} 
\altaffiltext{3} {Hubble Fellow}
\altaffiltext{4}{Canadian Institute for Theoretical Astrophysics, 60 St George
St, Toronto, ON M5S 3H8, Canada, email: pchang@cita.utoronto.ca}

\begin{abstract}
  We discuss the nature of rings that may exist around extrasolar planets.
Taking the general properties of rings around the gas giants in the Solar
System, we infer the likely properties of rings around exoplanets that reside
inside the ice line. Due to their proximity to their host star, rings around
such exoplanets must primarily consist of rocky materials. However, we find
that despite the higher densities of rock compared to ice, most of the
observed extrasolar planets with reliable radii measurements have sufficiently
large Roche radii to support rings. For the currently known transiting
extrasolar planets, Poynting-Robertson drag is not effective in significantly
altering the dynamics of individual ring particles over a time span of $10^8$
years provided that they exceed about 1 m in size. In addition, we show that
significantly smaller ring particles can exist in optically thick rings, for
which we find typical ring lifetimes ranging from a few times $10^6$ to a few
times $10^9$ years. Most interestingly, we find that many of the rings could
have nontrivial Laplacian planes due to the increased effects of the orbital
quadrupole caused by the exoplanets' proximity to their host star, allowing a
constraint on the $J_2$ of extrasolar planets from ring observations. This is
particular exciting, since a planet's $J_2$ reveals information about its
interior structure. Furthermore, measurements of an exoplanet's oblateness and
of its $J_2$, from warped rings, would together place limits on its spin
period. Based on the constraints that we have derived for extrasolar rings, we
anticipate that the best candidates for ring detections will come from transit
observations by the Kepler spacecraft of extrasolar planets with semi-major
axes $\sim 0.1$~AU and larger.
\end{abstract}


\section{Introduction}\label{sec:intro}

Ring systems exist around all of the giant planets in our Solar System. The
rings of the Saturnian system are the most prominent and consist mainly of
centimeter to meter sized icy bodies \citep{FN00}.  On the other hand,
Jupiter's rings are far more tenuous and consist of micron-size dust particles
\citep{Showalter+08}. Since rings are ubiquitous around giant planets in the
Solar System, they may also be common around extrasolar planets.

Although more than 500 extrasolar planets have been discovered to date, no
extrasolar satellites or ring systems have been detected yet. However, this
may change soon due to the unprecedented photometric accuracy of the Kepler
satellite \citep{BKB10} and due to the constantly improving precision and
increasing temporal baseline of ground based radial velocity surveys. Since
rings typically reside in the planet's equatorial plane, the required
photometric and spectroscopic precision for ring detection depends of the
planet's obliquity. The obliquity, $\theta_*$, refers here to the angle
between an extrasolar planet's spin axis and the normal of its orbital
plane. \citet{BF04} estimate that Saturn-like rings could be detected around
transiting extrasolar planets with a photometric precision of $(1-3)\times
10^{-4}$ and a 15 minutes time resolution as long as the ring is not viewed
close to edge-on (i.e., as long as $\theta_*$ is not $\ll 1$). This is within
the photometric accuracy that the Kepler spacecraft achieves for Sun-like and
brighter stars (http://keplergo.arc.nasa.gov/CalibrationSN.shtml). In
addition, rings around transiting extrasolar planets could also be identified
spectroscopically \citep{OTS09}. \citet{OTS09} showed that rings with
significant obliquities are detectable with currently achievable radial
velocity precision of 1 m/s, whereas rings with $\theta_* \ll 1$ would
typically require a radial velocity precision of 0.1 m/s or less, which is
still beyond the reach of radial velocity surveys.

A potential obstacle to detecting extrasolar rings may be
that most close in exoplanets could have low obliquities, which would make
their rings hard, if not impossible, to discover. The initial obliquities of
close in extrasolar planets with masses comparable to and bigger than Neptune
are likely to be large, since such planets are thought to have formed at
larger semi-major axes and have reached their current location by
planet-planet scattering, disk migration, or by Kozai oscillations with a
stellar companion or a combination of such processes
\citep[e.g.][]{LP79,Lin+96,Rasio+96,CF08,Wu+03,Wu+07}. Tides raised on the
exoplanet by its host star will, however, lead to damping of its obliquity. To
first order in $\theta_*$, the obliquity damping timescale for exoplanets with
small eccentricities is given by
\begin{equation}\label{e0}
t_{damp} = \theta_* \frac{dt}{d\theta_*} \sim \frac{2 \alpha_{P} Q_{P}}{3
k_P} \left(\frac{M_P}{M_{*}}\right) \left(\frac{a}{R_P}\right)^3 \Omega^{-1}
\end{equation}
where $k_P$ is the exoplanet's tidal Love number, $Q_P$ its tidal dissipation
function, $M_*$ is the stellar mass and $a$, $R_P$ and $M_P$ are the
semi-major axis, radius and mass of the extrasolar planet, respectively
\citep[e.g.][]{H81,L07}. $\alpha_P=I_P/M_PR_P^2 \leq 2/5$, where $I_P$ is the
exoplanet's moment of inertia and $\Omega$ is its orbital frequency
\footnote{We have used the moment of inertia of a constant density sphere $I_P
  = (2/5) M_p R_p^2$ for our estimates (i.e., $\alpha_P=2/5$). The actual
  moment of inertia of a planet should be somewhat smaller than this because
  it will be centrally concentrated, i.e. its density will increase towards
  its center.}. Since the synchronization timescale is comparable to the
  obliquity damping timescale, we assumed in equation (1) that the exoplanet's
  spin period is comparable to its orbital period. Evaluating equation
  (\ref{e0}) for a Jupiter like exoplanet around a Sun-like star and assuming
  $Q_P \sim 10^{6.5}$ \citep{J08} and $k_P \sim 3/2$ we find that $t_{damp}
  \gtrsim 10^8~\rm{years}$ and $t_{damp} \gtrsim 10^9~\rm{years}$ for
  semi-major axes greater than about 0.1~AU and 0.2~AU, respectively. We
  therefore expect most exoplanets with semi-major axes greater than a few
  tenths of an AU to have significant obliquities, allowing for ring
  detections. Although only a handful of transiting exoplanets are currently
  known with $a \gtrsim 0.1$~AU, the Kepler satellite is likely to fill in
  this parameter space in the near future. Furthermore, even for systems with
  $a \lesssim 0.1$~AU, stellar tides do not need to damp exoplanets'
  obliquities to zero, because for sufficiently high initial obliquities, the
  planets may settle into a high obliquity Cassini state
  \citep{Winn+05,Fabrycky+07b,L07}. In short, we expect most exoplanets with
  semi-major axes greater than a few tenths of an AU to have significant
  obliquities, allowing for ring detections, and note that systems with
  smaller semi-major axes could reside in high obliquity Cassini states rather
  than having their obliquities damped to zero.

In this paper, we investigate what types of ring systems could exist around
extrasolar planets with semi-major axes of about 1AU or less.  We focus on
these systems, which we coin ``warm Saturns``, since they fall within the
Kepler discovery space, which is limited to extrasolar planets with orbital
periods of about 1 year and less. We show that such extrasolar ring systems,
if they exist, will differ from those in our Solar System and examine the
different dynamical forces that play a role in shaping them. We show that the
presence of extrasolar rings, or the lack thereof, provides interesting
implications for ring formation theories and that the detection of extrasolar
rings will constrain the extrasolar planet's obliquity and in some cases also
its quadrupole moment. Measuring an exoplanet's quadrupole moment would be
especially exciting since it would allow us to probe its interior structure
\citep{Ragozzine+09}.

This paper is structured as follows. We start by determining the Roche radius
and ring composition in section 2.1 and examine the effect of
Poynting-Robertson drag on the ring particles in section 2.2. In section 2.3
and 2.4, we discuss the implications of the planet's proximity to its host
star on ring formation and ring orientation, respectively. Discussion and
conclusions follow in section 3.

\section{Properties of Planetary Rings}\label{sec:planetary rings}

In this section, we discuss the general properties of planetary rings from
ring studies in the Solar System.  We then extend these results to extrasolar
planets and discuss their implications.

\subsection{Roche Radius \& Ring Composition}

The existence and radial extent of planetary rings is determined by the tidal
field of the planet. To make this discussion more concrete, we focus on the
Saturnian system. Within the planet's Roche radius, a satellite cannot attain
hydrostatic equilibrium, which typically leads to mass loss and the disruption
of the satellite and the subsequent formation of rings \citep{R1847,C69}. In
particular, for a large, self-gravitating and synchronously rotating satellite
with a density $\rho$ the Roche radius, $R_{\rm{Roche}}$, is
\begin{equation}\label{eq:roche}
\frac {\rroche}{\Rp} = 2.45 \left(\frac{\rhop}{\rho}\right)^{1/3},
\end{equation}
where $\rhop = 3\Mp/4\pi\Rp^3$ is the average density of the
planet \citep{Murray_and_Dermott}. For icy particles that make up the
Saturnian system, the average density is $0.5-0.9\,{\rm g\,cm}^{-3}$, while
the density of Saturn is $\approx 0.7\,{\rm g\,cm}^{-3}$. Hence, from equation
(\ref{eq:roche}), we have that Saturn's ring system should extend out to
approximately twice Saturn's planetary radius, which is consistent with the
observed rings around Saturn.

\begin{figure}
  \plotone{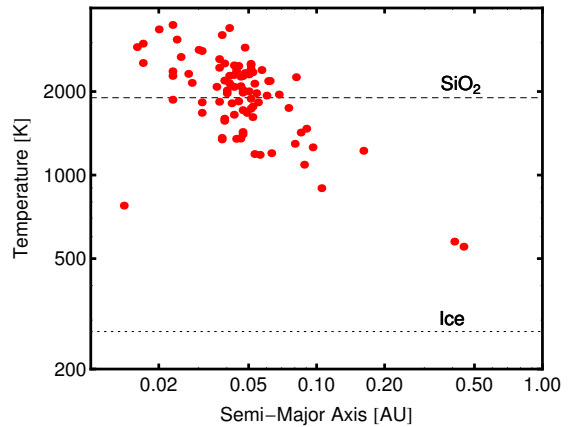}
  \caption{Equilibrium blackbody temperature for ring particles for known
    transiting extrasolar planets. The melting temperature of water ice
    (dotted line) and silicon dioxide ($\rm{SiO_2}$) (dashed line) are
    plotted for comparison. Exoplanet data is taken from \citet{W11}
    (http://exoplanets.org).}
   \label{f:temperature semi}
\end{figure}

The icy particles that make up Saturn's rings can exist at Saturn's orbital
radius because the local temperature is sufficiently low. However, for the
known extrasolar planets, the presence of ices is doubtful as most of the them
reside close to their parent star. In Figure \ref{f:temperature semi}, we plot
the equilibrium blackbody temperature for ring particles of known transiting
extrasolar planets. All of these planets have blackbody effective temperatures
well in excess of the melting temperature of water ice. There is a
considerable range in melting and sublimation temperatures for different
compositions of rock. For comparison we plot the melting temperature of
silicon dioxide ($\rm{SiO_2}$), which is a high-melting point solid. Comparing
the equilibrium blackbody temperatures for the currently known transiting
exoplanet with the melting temperature of $\rm{SiO_2}$ suggests that up to
about 35 extrasolar planets could harbor rings made of rocky material. The
blackbody equilibrium temperature shown in Figure \ref{f:temperature semi} was
calculated from the exoplanet's semi-major axis. Ring particles around
eccentric exoplanets may therefore reach maximum temperatures that exceed the
temperatures plotted in Figure \ref{f:temperature semi}.

\begin{figure}
  \plotone{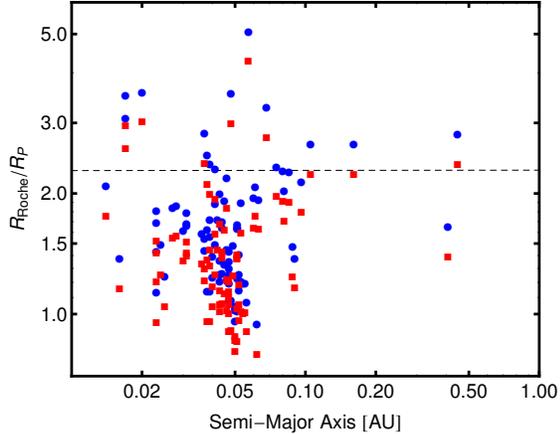}
  \caption{Roche radius, $\rroche$, of currently known transiting extrasolar
planets for a particle density of 3 (blue circles) and $5\,{\rm g\,cm}^{-3}$
(red squares). The dashed line corresponds to Saturn's Roche radius with a
mean density of $\rho_{\rm{P}}=0.7\,{\rm g\,cm}^{-3}$ and for icy ring
particles with a density of $\rho=1\,{\rm g\,cm}^{-3}$. It is clear from this
plot that a significant number of extrasolar planets have Roche radii that
allow for the existence of rings. The exoplanet data are taken from
\citet{W11} (http://exoplanets.org).}
   \label{f:ring radii}
\end{figure}

The density of rock varies between 2 and 5 ${\rm g\,cm}^{-3}$ depending on
composition, i.e., iron/nickel content, and porosity.  The higher density of
rock compared to ice implies that the resulting ring systems would be more
compact compared to icy ring systems (see equation (\ref{eq:roche})). Still a
substantial number of extrasolar planets could potentially support rings.  We
show this in Figure \ref{f:ring radii} where we plot the Roche radius of
currently known transiting extrasolar planets for ring particle densities of 3
(blue circles) and $5\,{\rm g\,cm}^{-3}$ (red squares). We also plot $\rroche$
for Saturn with a mean density of $\rho_{\rm{P}} = 0.7\,{\rm g\,cm}^{-3}$ and
water ice ring particles with $\rho=1\,{\rm g\,cm}^{-3}$. It is clear from
this plot that a number of extrasolar planets can support rings made of rocky
material. Indeed 86(76) of the 88 planets with reliable radius measurements
can support rocky rings with a material density of $\rho = 3(5)\gcc$, i.e.,
$\rroche/\Rp > 1$.  Of these planets, 21(12) or $ 24\%$ ($14\%$) can support
sizable rings, i.e., $\rroche/\Rp > 2$.
 
\subsection{Poynting-Robertson Drag} 

Having calculated the equilibrium temperatures and the sizes of the Roche
radii of extrasolar planets, we now turn to examining the ring lifetimes due
to Poynting-Robertson drag. In the Solar System, Poynting-Robertson drag is
not important for Saturn's rings, but it does drive the evolution of particles
in Jupiter's rings \citep{Burns+99,Showalter+08}. Because of the larger
stellar insolation of warm Saturns, Poynting-Robertson drag is significant
even for large ring particles as we show below.

\begin{figure}
  \plotone{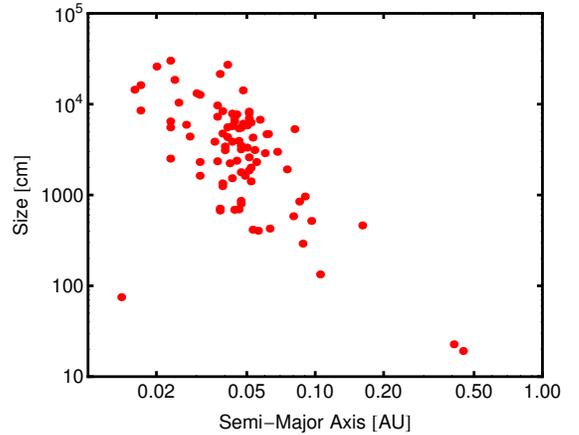}
  \caption{The smallest ring particle size for which $t_{\rm
  PR}>10^8~\rm{years}$ of known transiting extrasolar planets. The
  Poynting-Robertson timescale, $t_{\rm PR}$, was evaluated assuming $Q_{\rm
  PR} \sim 0.5$ and $i \sim 45^{\circ}$. The exoplanet data used in this
  calculation are from \citet{W11} (http://exoplanets.org).}
   \label{f:PRSize}
\end{figure}

The orbital decay time, $t_{\rm PR}$, of a circumplanetary ring particle with
radius, $s$, due to Poynting-Robertson drag is given by
\begin{equation}\label{e:2}
t_{\rm PR} \sim \frac{8 \rho s c^2}{3(L/4\pi a^2) Q_{\rm PR}(5+\cos^2(i))}
\end{equation}
where $c$ is the speed of light, $L$ the stellar luminosity, $i$ the
inclination of the ring plane with respect to the orbital plane of the
extrasolar planet and $Q_{\rm PR}$ is the radiation pressure efficency factor
\citep{B79}. If the orbital evolution of each ring particle can be considered
independently and if mutual shadowing of ring particles can be neglected, then
equation (\ref{e:2}) yields the ring particle lifetime due to
Poynting-Robertson drag. Figure \ref{f:PRSize} shows the smallest ring
particles that can survive over $10^8$ years in known transiting extrasolar
planet systems due to Poynting-Robertson drag provided that each ring particle
evolves independently. From equation (\ref{e:2}) we see that $t_{\rm{PR}}$ is
considerably shorter for small ring particles, suggesting a considerable
amount of ring spreading due to Poynting-Robertson drag. It is, however,
likely that the evolution of individual ring particles are coupled to each
other by frequent collisions, in which case the size dependence of
$t_{\rm{PR}}$ is averaged out.

\begin{figure}
  \plotone{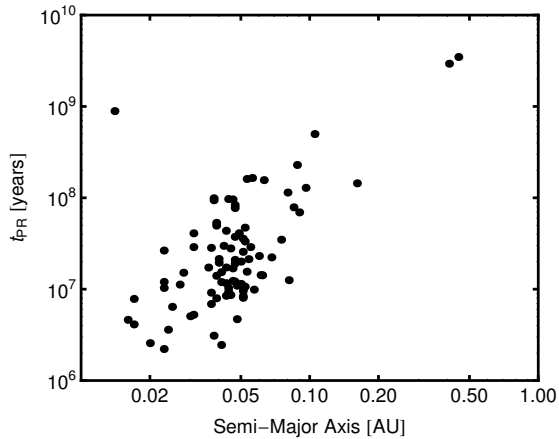}
 \caption{Ring lifetimes due to Poynting-Robertson drag assuming optically
 thick planetary rings around known transiting extrasolar planets. The
 Poynting-Robertson timescale, $t_{\rm PR}$, was evaluated assuming $Q_{\rm
 PR} \sim 0.5$, $i \sim 45^{\circ}$ and $\Sigma \sim 400~\rm{g~cm^{-2}}$.}
\label{f:PRtime}
\end{figure}

On the other hand, if the ring is optically thick, then the Poynting-Robertson
drag timescale depends on the ring mass surface density instead of the sizes
of individual ring particles (see equation (\ref{e:3})). For an optically
thick ring the maximum surface area that is exposed to stellar irradiation is
$\pi \sin i (R_{out}^2-R_{in}^2)$, where $R_{out}$ and $R_{in}$ are the outer
and inner ring radii, respectively. Averaging over the orbit of the planet
around the star holding the ring orientation fixed, we find that the average
surface area exposed to the host star is $2 \sin i (R_{out}^2-R_{in}^2)$. This
yields an orbital decay time due to Poynting-Robertson drag given by
\begin{equation}\label{e:3}
t_{\rm PR} \sim \frac{\pi c^2 \Sigma}{\sin i(L/4\pi a^2) Q_{\rm
PR}(5+\cos^2(i))}
\end{equation}
where $\Sigma$ is the mass surface density of the ring. Figure \ref{f:PRtime}
shows the ring lifetimes for known extrasolar planets for ring mass surface
densities comparable to Saturn's B-ring (i.e., $\Sigma \sim
400~\rm{g~cm^{-2}}$) \citep{R10}. Since $t_{\rm PR}$ scales as $\Sigma$, we
note here, that the ring lifetimes could be significantly longer for ring
systems with ring mass surface densities larger than that of
Saturn. Furthermore, the ring lifetimes in Figure \ref{f:PRtime} were
calculated for an inclinations of $45^{\circ}$, rings with smaller
inclinations than this reference value would have longer lifetimes, since a
smaller effective ring surface area would be exposed to the stellar radiation
from the host star.

Figure \ref{f:PRSize} suggests that ring particles, if they evolve
individually, need to be about a meter and larger for the existence of
long-lived rings (i.e. $t >10^8$~years) around the currently known transiting
extrasolar planets. If however the ring is optically thick, then the
Poynting-Robertson drag timescale depends only on the mass surface density of
the ring and significantly smaller ring particles can survive over long
periods. We note here that the actual ring lifetimes could be shortened due to
ring spreading caused by collisions between ring particles, differential
precession and/or Poynting-Robertson drag \citep{GT79,GT82}.

\subsection{Formation} 
Extrasolar planets with masses comparable to Neptune and larger on short
period orbits probably did not form in situ but reached their current location
by either planet-planet scattering, migration or by Kozai oscillations with a
stellar companion \citep[e.g.][]{LP79,Lin+96,Rasio+96,CF08,Wu+03,Wu+07}. If
such planets originally had icy rings, then these rings will have been
sublimated by the time they arrived at their current semi-major axes. This
suggests that, if extrasolar rings are discovered around such planets, that
they probably formed close to their current semi-major axes, which may have
interesting implications for ring formation.

\begin{figure}
  \plotone{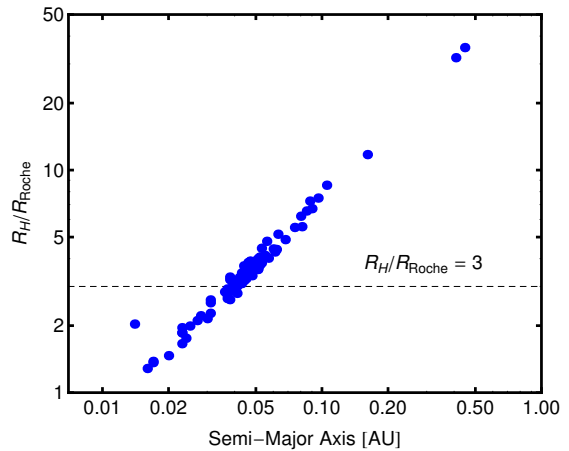}
 \caption{Ratio of the Hill radius, $\RH$, to the Roche radius, $\rroche$, of
 known extrasolar planets. The dashed line corresponds to $\RH/\rroche=3$. The
 exoplanet data are taken from \citet{W11} (http://exoplanets.org)}
\label{f:RHRR}
\end{figure}

The Hill radius, $\RH$, denotes the distance from a planet at which the tidal
forces from its host star and the gravitational forces from the planet both
acting on a test particle, are in equilibrium. It is given by
\begin{equation}
\RH = a \left( \frac{\Mp}{3M_{*}}\right)^{1/3}.
\end{equation}
In our Solar Systems planetary rings typically reside well inside the Hill
sphere of their respective hosts. This is because in our Solar System
$\rroche<< \RH$. For some extrasolar planets however, $\rroche \sim \RH$, due
to the proximity to their host stars. Since the outer regions of the Hill
sphere are unstable \citep[e.g.][]{H69,H70,I79,HB91,SS08}, no planetary rings
can exist there. The permitted range within which bound stable orbits, and
therefore rings, can exist depends on the inclination of the ring particle's
orbit. Retrograde orbits are in general more stable than prograde orbits. For
example, coplanar prograde orbits are stable within about $\RH/3$ whereas
coplanar retrograde orbits are stable within about $2\RH/3$
\citep[e.g.][]{H69,VW01}. The unstable outer parts of the Hill sphere could
have interesting implications for ring formation scenarios. If rings are
formed by a larger body that sheds mass as it comes within the Roche radius of
a given extrasolar planet, then extrasolar planets with $\rroche \sim \RH$ are
at a disadvantage, since, due to the lack of bound, stable orbits in the outer
parts of the Hill sphere, mass shed in this region will be lost from the
system and will therefore not be available for ring formation. Therefore for
prograde rings, an extrasolar planet with $\rroche < \RH/3$ may be a better
candidate for hosting rings than one with $\rroche > \RH/3$ (see Figure
\ref{f:RHRR}).

\subsection{Ring Orientation}\label{sec:orientation}

The ring orientation for some of these warm Saturns may not be trivial, since
it is determined by the competing forces of the planet's bulge and the stellar
tide.  Because the ratio of these forces varies as a function of the ring's
distance from the planet, $r$, the ring's orientation follows the planet's
equator at small $r$ and follows the orbital plane at large $r$.

\begin{figure}
  \plotone{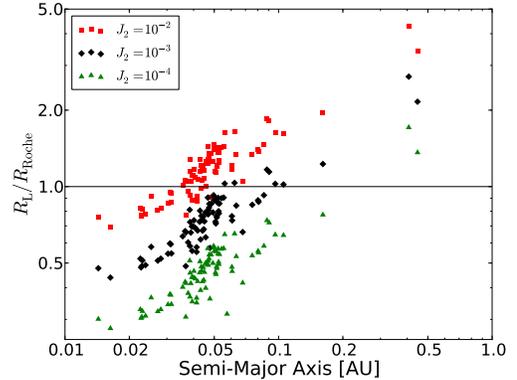}
 \caption{Ratio of the Laplace radius to the Roche radius for ring material
 with a density of $3\gcc$ and $J_2=10^{-4}-10^{-2}$. The solid line marks
 where $\rL=\rroche$. Above this line, the rings will mostly lie in the plane
 defined by the planet's equator. Whereas below this line, the rings will
 undergo a transition from lying in the planet's equatorial plane at small $r$
 to lying in the orbital plane at large $r$. The exoplanet data used in this
 calculation are from \citet{W11} (http://exoplanets.org).}
\label{f:aL}
\end{figure}

The combined effects of the planet's oblateness and the stellar tide in
determining the ring orientation was first recognized by \citet{Laplace1805}.
Here we use the more recent discussion of \citet[hereafter TTN]{Tremaine+09}.
Because the strength of planetary oblateness and the stellar tide scale
differently with the planet ring separation, the ring orientation varies as a
function of $r$.  The plane that this defines is known as the Laplace
plane. To estimate the magnitude of this effect, we first note that the
strength of the quadrupole potential arising from the planet's bulge is (TTN)
\begin{equation}\label{eq:planet quadrupole}
 \Phi_p = \frac{GM_pJ_2 \Rp^2}{r^3}P_2(\cos\theta),
\end{equation}
where $\theta$ is the polar angle from the rotation axis of the planet, $J_2$ is the quadrupole gravitational harmonic, and $P_2$ is a Legendre polynomial.
The quadrupole potential arising from the star is 
\begin{equation}\label{eq:stellar quadrupole}
 \Phi_* = \frac{GM_* r^2}{2a^3(1-e_*)^{3/2}}P_2(\cos\theta_*),
\end{equation}
where $e_*$ is the extrasolar planet's eccentricity. Equating equations
(\ref{eq:planet quadrupole}) and (\ref{eq:stellar quadrupole}) and ignoring
the $P_2$ terms,\footnote{Another way of looking at this is to assume $\theta$
and $\theta_*$ are $\approx \pi/2$ so that $P_2(\theta,\theta_* ) \approx
1/2$} we estimate what is known as the Laplace radius, $\rL$:
\begin{equation}\label{eq:laplace radius}
 \rL^5 = 2 J_2 \Rp^2 a^3 \left(1-e_*\right)^{3/2}\frac {M_p} {M_*}
\end{equation}
This simple order of magnitude estimate agrees with the exact calculation of
TTN.\footnote{TTN lacks our factor of 2, which is instead absorbed into their equation for the Laplace equilibria, i.e., their equation (23).}
Numerically this gives
\begin{eqnarray}
\frac{\rL}{\Rp} &\approx & 2.9 \left(\frac{J_2}{0.01}\right)^{1/5} \left(\frac{(a/0.1\,{\rm AU})}{(\Rp/R_J)}\right)^{3/5}\nonumber\\
&&\left(\frac {M_p/M_*}{0.001}\right)^{1/5}\left(1-e_*\right)^{3/10},
\end{eqnarray}
where $R_J = 71492$ km is the radius of Jupiter. To determine if
the rings will lie in the equatorial plane of the planet or in the planet's
orbital plane around the host star, we take the ratio of $\rL$ and $\rroche$
\begin{eqnarray}
\frac{\rL}{\rroche} &\approx& 0.75\left(\frac {J_2}{0.01}\right)^{1/5} \left(\frac{M_p/M_*}{0.001}\right)^{-2/15}\left(\frac{\Rp}{R_{\rm J}}\right)^{2/5}\nonumber \\
&& \times\left(\frac{a}{0.1\,{\rm AU}}\right)^{3/5} \left(\frac{\rho}{3\,{\rm g\,cm}^{-3}}\right)^{1/3}.
\end{eqnarray}
In Figure \ref{f:aL}, we plot this ratio for three different values of $J_2$,
ranging from of $10^{-4}$ to $10^{-2}$. For reference, we note that the giant
planets in the Solar System have $J_2$'s that vary from $\approx 0.003$ for
Uranus and Neptune to $\approx 0.01$ for Jupiter and Saturn.  $\rL = \rroche$
is denoted by the solid line in Figure \ref{f:aL}. Above this line,
$\rL>\rroche$ and the rings will mostly lie in the plane defined by the
planet's equator. Below this line, $\rL < \rroche$ and the rings will undergo
a transition from lying in the planet's equatorial plane at small $r$ to lying
in the orbital plane at large $r$. From Figure \ref{f:aL}, it is clear that
the fraction of planets with nontrivial Laplacian planes varies with
$J_2$. For $J_2 \lesssim 10^{-3}$, most ringed extrasolar planets fall below
this line and thus have warped rings such that their rings will lie in the
planet's equatorial plane inside of $\rL$, but coincide with the orbital plane
outside of $\rL$.  On the other hand, for $J_2 = 10^{-2}$, most planets have
rings that lie in the plane defined by the exoplanet's equator, much like the
planetary rings in the Solar System.

The observational signature of warped rings is especially interesting as it
provides a means by which the planet's $J_2$ can be measured directly. Present
constraints on $J_2$ are inferred from transit measurements of the planet's
oblateness \citep{Carter+10a}. The inferred $J_2$ from such oblateness
measurements is however model dependent. Since warped rings provide a direct
constraint on the planet's $J_2$, which in turn relates to the three moments
of inertia about the principle axes, the planet's internal structure can be
probed \citep{Ragozzine+09}. Furthermore, measurements of an exoplanet's $J_2$
and of its oblateness would together constrain its spin period. This method
was successfully applied in the past to determine the rotation period of Uranus
\citep{DE79,EF81}.

\section{Discussion and Conclusions}
We examined the nature of rings that could exist around extrasolar planets
that have orbital periods of about one year or less. Such systems are ideal
targets for the Kepler satellite, whose photometric precision will be able to
identify Saturn-like rings around extrasolar planets that are transiting
Sun-like stars \citep{BF04} (http:keplergo.arc.nasa.gov/CalibrationSN.shtml).

We have shown that most currently known transiting extrasolar planets are too
close to their parent star to support icy rings but that a significant
fraction of them could harbor ring particles made of rock or silicates. We
calculated the Roche radius for currently known transiting extrasolar planets
and compared it with that of Saturn. Most currently known transiting
extrasolar planets have Roche radii large enough to support rings and 12 to 21
of them, have Roche radii that are comparable to or larger than Saturn's Roche
radius, suggesting that such extrasolar planets could harbor sizable rings. In
addition, we examined the ring lifetime due to Poynting-Robertson drag. For
optically thick rings and a ring mass surface density similar to that of
Saturn's B-ring, we find ring lifetimes typically range from a few times
$10^6$ to a few times $10^9$ years. We note here that the actual ring
lifetimes could be shortened due to ring spreading \citep{GT79,GT82}. Finally,
we showed that, in contrast to the rings in the Solar System, some of these
extrasolar rings may be warped because of the competing effects of planetary
and stellar tide. Observations of warped rings would provide a direct
measurement of the planet's $J_2$.  This is particular exciting, since a
planet's $J_2$ reveals information about its interior structure
\citep{Ragozzine+09}. Previous constraints on the $J_2$ of extrasolar planets
are model dependent as they are derived from the exoplanet's oblateness, which
is determined from transit light curves
\citep{Seager+02,Barnes+03,Carter+10b,Leconte+11}. For example,
\citet{Carter+10a} recently placed constraints on the $J_2$ of HD189733b to be
$<0.068$. Furthermore, measurements of an exoplanet's $J_2$ from warped rings
and of its oblateness, would together place limits on its spin period.

Close in extrasolar planets with masses comparable to Neptune and larger are
generally thought to have formed outside the ice line and to have reached
their current location by either planet-planet scattering, disk migration, or
by Kozai oscillations with a stellar companion
\citep[e.g.][]{LP79,Lin+96,Rasio+96,CF08,Wu+03,Wu+07}. If such planets
originally formed with icy rings, such rings would have been sublimated by the
time they arrived at their current semi-major axes. This suggests that, if
extrasolar rings are discovered, they probably formed close to their current
location. We showed that due to the proximity to their host stars, $\rroche
\sim \RH$ for some extrasolar planets. This has interesting implications for
ring formation, since orbits in the outer regions of the Hill sphere are
chaotic and often unbound, which makes this region unsuitable for harboring
rings. If rings are formed by a larger body that sheds mass as it comes within
the Roche radius of a given extrasolar planet, then extrasolar planets with
$\rroche \sim \RH$ are at a disadvantage, since, due to the lack of bound,
stable orbits in the outer parts of the Hill sphere, mass shed in this
region will be lost from the system and will therefore not be available for
ring formation. Therefore an extrasolar planet with $\rroche \ll \RH$ may be a
better candidate for hosting rings than one with $\rroche \sim \RH$.

The observation of extrasolar rings can offer interesting constraints on the
obliquity distribution of extrasolar planets. These insights, in turn, may
help to differentiate between various proposed mechanisms by which these warm
Saturns were transported to their current location. For example, if
planet-planet scattering \citep{Rasio+96,CF08} and/or Kozai oscillations with a
stellar companion \citep{Wu+03,Wu+07} are responsible for the observed small
semi-major axes of many exoplanets, then the extrasolar planet's obliquities
should be large. If on the other hand, migration in a gaseous disk is
primarily responsible for the current location of close in extrasolar planets
then their obliquities are likely to be small \citep{LP79,Lin+96}. The
obliquity distribution of extrasolar planets therefore provides a valuable
probe for differentiating between these proposed planet formation scenarios.
Recent measurements of the Rossiter-McLaughlin effect find a strong
misalignment between the normal of the orbital plane and the stellar spin axis
for some exoplanets \citep[e.g.][]{Winn+10}, which is consistent with
expectations from planet-planet scattering and Kozai oscillations.

Furthermore, for sufficiently small semi-major axes, stellar tides will act to
damp the exoplanet's obliquity \citep{Goldreich+70,H81}, while preserving the
spin axis orientation at large semi-major axes (see equation \ref{e0}). Hence,
one may expect to see a transition from small to larger obliquities as a
function of semi-major axis. Observations of such a transition in the
obliquity distribution of exoplanets could in principle be used to infer the
tidal dissipation function, $Q_P$. However, the actual obliquity evolution may
be complicated by interactions with other planets in the system, as
\citet{Laskar+93} have shown to be important for the terrestial planets in the
Solar System. In addition, stellar tides do not need to damp the planet's
obliquity to zero, because for sufficiently high initial obliquities, the
planets may settle into a high obliquity Cassini state
\citep{Winn+05,Fabrycky+07b}.

In summary, given the various requirements for harboring rings and the fact
that rings are most easily discovered around exoplanets with significant
obliquities \citep{BF04,OTS09}, we conclude that the majority of the currently
known transiting extrasolar planets examined here are not ideal candidates for
ring detections, since most of them are too close to their parent star. We
find, however, no compelling reason arguing against the detection of rings
around exoplanets with semi-major axes $\gtrsim 0.1~\rm{AU}$, which is very
exciting since the Kepler satellite will probe this parameter space.

\acknowledgements{We thank Peter Goldreich for insightful discussions.  We
thank the anonymous referee for useful comments that helped to improve this
manuscript. This research has made use of the Exoplanet Orbit Database and
Exoplanet Data Explorer at exoplanets.org. HS is supported by NASA through
Hubble Fellowship Grant \# HST-HF-51281.01-A awarded by the Space Telescope
Science Institute, which is operated by the Association of Universities for
Research in Astronomy, Inc., for NASA, under contact NAS 5-26555. PC is
supported by the Canadian Institute for Theoretical Astrophysics.}

\bibliographystyle{apj} 

\begin{thebibliography}{44}
\expandafter\ifx\csname natexlab\endcsname\relax\def\natexlab#1{#1}\fi

\bibitem[{{Barnes} \& {Fortney}(2003)}]{Barnes+03}
{Barnes}, J.~W. \& {Fortney}, J.~J. 2003, \apj, 588, 545

\bibitem[{{Barnes} \& {Fortney}(2004)}]{BF04}
---. 2004, \apj, 616, 1193

\bibitem[{{Borucki} {et~al.}(2010){Borucki}, {Koch}, {Basri}, {Batalha},
  {Brown}, {Caldwell}, {Caldwell}, {Christensen-Dalsgaard}, {Cochran},
  {DeVore}, {Dunham}, {Dupree}, {Gautier}, {Geary}, {Gilliland}, {Gould},
  {Howell}, {Jenkins}, {Kondo}, {Latham}, {Marcy}, {Meibom}, {Kjeldsen},
  {Lissauer}, {Monet}, {Morrison}, {Sasselov}, {Tarter}, {Boss}, {Brownlee},
  {Owen}, {Buzasi}, {Charbonneau}, {Doyle}, {Fortney}, {Ford}, {Holman},
  {Seager}, {Steffen}, {Welsh}, {Rowe}, {Anderson}, {Buchhave}, {Ciardi},
  {Walkowicz}, {Sherry}, {Horch}, {Isaacson}, {Everett}, {Fischer}, {Torres},
  {Johnson}, {Endl}, {MacQueen}, {Bryson}, {Dotson}, {Haas}, {Kolodziejczak},
  {Van Cleve}, {Chandrasekaran}, {Twicken}, {Quintana}, {Clarke}, {Allen},
  {Li}, {Wu}, {Tenenbaum}, {Verner}, {Bruhweiler}, {Barnes}, \& {Prsa}}]{BKB10}
{Borucki}, W.~J., {Koch}, D., {Basri}, G., {Batalha}, N., {Brown}, T.,
  {Caldwell}, D., {Caldwell}, J., {Christensen-Dalsgaard}, J., {Cochran},
  W.~D., {DeVore}, E., {Dunham}, E.~W., {Dupree}, A.~K., {Gautier}, T.~N.,
  {Geary}, J.~C., {Gilliland}, R., {Gould}, A., {Howell}, S.~B., {Jenkins},
  J.~M., {Kondo}, Y., {Latham}, D.~W., {Marcy}, G.~W., {Meibom}, S.,
  {Kjeldsen}, H., {Lissauer}, J.~J., {Monet}, D.~G., {Morrison}, D.,
  {Sasselov}, D., {Tarter}, J., {Boss}, A., {Brownlee}, D., {Owen}, T.,
  {Buzasi}, D., {Charbonneau}, D., {Doyle}, L., {Fortney}, J., {Ford}, E.~B.,
  {Holman}, M.~J., {Seager}, S., {Steffen}, J.~H., {Welsh}, W.~F., {Rowe}, J.,
  {Anderson}, H., {Buchhave}, L., {Ciardi}, D., {Walkowicz}, L., {Sherry}, W.,
  {Horch}, E., {Isaacson}, H., {Everett}, M.~E., {Fischer}, D., {Torres}, G.,
  {Johnson}, J.~A., {Endl}, M., {MacQueen}, P., {Bryson}, S.~T., {Dotson}, J.,
  {Haas}, M., {Kolodziejczak}, J., {Van Cleve}, J., {Chandrasekaran}, H.,
  {Twicken}, J.~D., {Quintana}, E.~V., {Clarke}, B.~D., {Allen}, C., {Li}, J.,
  {Wu}, H., {Tenenbaum}, P., {Verner}, E., {Bruhweiler}, F., {Barnes}, J., \&
  {Prsa}, A. 2010, Science, 327, 977

\bibitem[{{Burns} {et~al.}(1979){Burns}, {Lamy}, \& {Soter}}]{B79}
{Burns}, J.~A., {Lamy}, P.~L., \& {Soter}, S. 1979, \icarus, 40, 1

\bibitem[{{Burns} {et~al.}(1999){Burns}, {Showalter}, {Hamilton}, {Nicholson},
  {de Pater}, {Ockert-Bell}, \& {Thomas}}]{Burns+99}
{Burns}, J.~A., {Showalter}, M.~R., {Hamilton}, D.~P., {Nicholson}, P.~D., {de
  Pater}, I., {Ockert-Bell}, M.~E., \& {Thomas}, P.~C. 1999, Science, 284, 1146

\bibitem[{{Carter} \& {Winn}(2010{\natexlab{a}})}]{Carter+10a}
{Carter}, J.~A. \& {Winn}, J.~N. 2010{\natexlab{a}}, \apj, 709, 1219

\bibitem[{{Carter} \& {Winn}(2010{\natexlab{b}})}]{Carter+10b}
---. 2010{\natexlab{b}}, \apj, 716, 850

\bibitem[{{Chandrasekhar}(1969)}]{C69}
{Chandrasekhar}, S. 1969, {Ellipsoidal figures of equilibrium}, ed.
  {Chandrasekhar, S.}

\bibitem[{{Chatterjee} {et~al.}(2008){Chatterjee}, {Ford}, {Matsumura}, \&
  {Rasio}}]{CF08}
{Chatterjee}, S., {Ford}, E.~B., {Matsumura}, S., \& {Rasio}, F.~A. 2008, \apj,
  686, 580

\bibitem[{{Dunham} \& {Elliot}(1979)}]{DE79}
{Dunham}, E. \& {Elliot}, J.~L. 1979, in Bulletin of the American Astronomical
  Society, Vol.~11, Bulletin of the American Astronomical Society, 568--+

\bibitem[{{Elliot} {et~al.}(1981){Elliot}, {French}, {Frogel}, {Elias}, {Mink},
  \& {Liller}}]{EF81}
{Elliot}, J.~L., {French}, R.~G., {Frogel}, J.~A., {Elias}, J.~H., {Mink},
  D.~J., \& {Liller}, W. 1981, \aj, 86, 444

\bibitem[{{Fabrycky} {et~al.}(2007){Fabrycky}, {Johnson}, \&
  {Goodman}}]{Fabrycky+07b}
{Fabrycky}, D.~C., {Johnson}, E.~T., \& {Goodman}, J. 2007, \apj, 665, 754

\bibitem[{{French} \& {Nicholson}(2000)}]{FN00}
{French}, R.~G. \& {Nicholson}, P.~D. 2000, \icarus, 145, 502

\bibitem[{{Goldreich} \& {Peale}(1970)}]{Goldreich+70}
{Goldreich}, P. \& {Peale}, S.~J. 1970, \aj, 75, 273

\bibitem[{{Goldreich} \& {Tremaine}(1979)}]{GT79}
{Goldreich}, P. \& {Tremaine}, S. 1979, \nat, 277, 97

\bibitem[{{Goldreich} \& {Tremaine}(1982)}]{GT82}
---. 1982, \araa, 20, 249

\bibitem[{{Hamilton} \& {Burns}(1991)}]{HB91}
{Hamilton}, D.~P. \& {Burns}, J.~A. 1991, Icarus, 92, 118

\bibitem[{{Henon}(1969)}]{H69}
{Henon}, M. 1969, \aap, 1, 223

\bibitem[{{Henon}(1970)}]{H70}
---. 1970, \aap, 9, 24

\bibitem[{{Hut}(1981)}]{H81}
{Hut}, P. 1981, \aap, 99, 126

\bibitem[{{Innanen}(1979)}]{I79}
{Innanen}, K.~A. 1979, \aj, 84, 960

\bibitem[{{Jackson} {et~al.}(2008){Jackson}, {Greenberg}, \& {Barnes}}]{J08}
{Jackson}, B., {Greenberg}, R., \& {Barnes}, R. 2008, \apj, 678, 1396

\bibitem[{Laplace(1805)}]{Laplace1805}
Laplace, P.~S. 1805, { M\'ecaniqe c\'eleste Volume 4, Book 8, (Paris:
  Courcier)}

\bibitem[{{Laskar} \& {Robutel}(1993)}]{Laskar+93}
{Laskar}, J. \& {Robutel}, P. 1993, \nat, 361, 608

\bibitem[{{Leconte} {et~al.}(2011){Leconte}, {Lai}, \& {Chabrier}}]{Leconte+11}
{Leconte}, J., {Lai}, D., \& {Chabrier}, G. 2011, ArXiv e-prints

\bibitem[{{Levrard} {et~al.}(2007){Levrard}, {Correia}, {Chabrier}, {Baraffe},
  {Selsis}, \& {Laskar}}]{L07}
{Levrard}, B., {Correia}, A.~C.~M., {Chabrier}, G., {Baraffe}, I., {Selsis},
  F., \& {Laskar}, J. 2007, \aap, 462, L5

\bibitem[{{Lin} {et~al.}(1996){Lin}, {Bodenheimer}, \& {Richardson}}]{Lin+96}
{Lin}, D.~N.~C., {Bodenheimer}, P., \& {Richardson}, D.~C. 1996, \nat, 380, 606

\bibitem[{{Lin} \& {Papaloizou}(1979)}]{LP79}
{Lin}, D.~N.~C. \& {Papaloizou}, J. 1979, \mnras, 186, 799

\bibitem[{{Murray} \& {Dermott}(2000)}]{Murray_and_Dermott}
{Murray}, C.~D. \& {Dermott}, S.~F. 2000, {Solar System Dynamics}

\bibitem[{{Ohta} {et~al.}(2009){Ohta}, {Taruya}, \& {Suto}}]{OTS09}
{Ohta}, Y., {Taruya}, A., \& {Suto}, Y. 2009, \apj, 690, 1

\bibitem[{{Ragozzine} \& {Wolf}(2009)}]{Ragozzine+09}
{Ragozzine}, D. \& {Wolf}, A.~S. 2009, \apj, 698, 1778

\bibitem[{{Rasio} \& {Ford}(1996)}]{Rasio+96}
{Rasio}, F.~A. \& {Ford}, E.~B. 1996, Science, 274, 954

\bibitem[{{Robbins} {et~al.}(2010){Robbins}, {Stewart}, {Lewis}, {Colwell}, \&
  {Srem{\v c}evi{\'c}}}]{R10}
{Robbins}, S.~J., {Stewart}, G.~R., {Lewis}, M.~C., {Colwell}, J.~E., \&
  {Srem{\v c}evi{\'c}}, M. 2010, \icarus, 206, 431

\bibitem[{{Roche}(1847)}]{R1847}
{Roche}, R.~A. 1847, Mem. de la Section des Sciences, 1, 243

\bibitem[{{Schlichting} \& {Sari}(2008)}]{SS08}
{Schlichting}, H.~E. \& {Sari}, R. 2008, \apj, 686, 741

\bibitem[{{Seager} \& {Hui}(2002)}]{Seager+02}
{Seager}, S. \& {Hui}, L. 2002, \apj, 574, 1004

\bibitem[{{Showalter} {et~al.}(2008){Showalter}, {de Pater}, {Verbanac},
  {Hamilton}, \& {Burns}}]{Showalter+08}
{Showalter}, M.~R., {de Pater}, I., {Verbanac}, G., {Hamilton}, D.~P., \&
  {Burns}, J.~A. 2008, \icarus, 195, 361

\bibitem[{{Tremaine} {et~al.}(2009){Tremaine}, {Touma}, \&
  {Namouni}}]{Tremaine+09}
{Tremaine}, S., {Touma}, J., \& {Namouni}, F. 2009, \aj, 137, 3706

\bibitem[{{Vieira Neto} \& {Winter}(2001)}]{VW01}
{Vieira Neto}, E. \& {Winter}, O.~C. 2001, \aj, 122, 440

\bibitem[{{Winn} {et~al.}(2010){Winn}, {Fabrycky}, {Albrecht}, \&
  {Johnson}}]{Winn+10}
{Winn}, J.~N., {Fabrycky}, D., {Albrecht}, S., \& {Johnson}, J.~A. 2010, \apjl,
  718, L145

\bibitem[{{Winn} \& {Holman}(2005)}]{Winn+05}
{Winn}, J.~N. \& {Holman}, M.~J. 2005, \apjl, 628, L159

\bibitem[{{Wright} {et~al.}(2010){Wright}, {Fakhouri}, {Marcy}, {Han}, {Feng},
  {Johnson}, {Howard}, {Fischer}, {Valenti}, {Anderson}, \& {Piskunov}}]{W11}
{Wright}, J.~T., {Fakhouri}, O., {Marcy}, G.~W., {Han}, E., {Feng}, Y.,
  {Johnson}, J.~A., {Howard}, A.~W., {Fischer}, D.~A., {Valenti}, J.~A.,
  {Anderson}, J., \& {Piskunov}, N. 2010, ArXiv e-prints

\bibitem[{{Wu} \& {Murray}(2003)}]{Wu+03}
{Wu}, Y. \& {Murray}, N. 2003, \apj, 589, 605

\bibitem[{{Wu} {et~al.}(2007){Wu}, {Murray}, \& {Ramsahai}}]{Wu+07}
{Wu}, Y., {Murray}, N.~W., \& {Ramsahai}, J.~M. 2007, \apj, 670, 820

\end{thebibliography}

\end{document}